\documentclass[aps,prl,superscriptaddress,showpacs,reprint]{revtex4-1}

\usepackage[utf8]{inputenc}
\usepackage{amsmath,amsfonts,amssymb}
\usepackage{pifont}
\usepackage{rotating}
\usepackage{enumerate}
\usepackage{graphicx}    
\usepackage{epstopdf}
\usepackage{color}
\usepackage{xcolor}
\usepackage{ulem}
\usepackage{soul} 
\usepackage{subfigure}
\usepackage[export]{adjustbox}

\usepackage{hyperref}
\definecolor{dark-red}{rgb}{0.9,0.15,0.15}
\definecolor{dark-blue}{rgb}{0.15,0.15,0.4}
\definecolor{medium-blue}{rgb}{0,0,0.5}
\hypersetup{
    colorlinks={true},
     linkcolor={medium-blue},
    citecolor={blue},
     urlcolor={medium-blue}
}

\begin{document}

\title{Bipolar Magnetic Semiconducting Behavior in VNbRuAl: A New Spintronic Material for Spin Filters}

\author{Jadupati Nag}
\affiliation{Department of Physics, Indian Institute of Technology Bombay, Mumbai 400076, India}

\author{Deepika Rani}
\thanks{These two authors have contributed equally to this work}
\affiliation{Department of Physics, Indian Institute of Technology Bombay, Mumbai 400076, India}

\author{Jiban Kangsabanik}
\thanks{These two authors have contributed equally to this work}
\affiliation{Department of Physics, Indian Institute of Technology Bombay, Mumbai 400076, India}
\affiliation{Computational Atomic-Scale Materials Design (CAMD), Department of Physics, Technical University of Denmark (DTU), Lyngby 2800 Kgs, Denmark}

\author{P. D. Babu}
\affiliation{UGC-DAE Consortium for Scientific Research, Mumbai Centre, BARC Campus, Mumbai 400085, India}

\author{K. G. Suresh}
\email{suresh@phy.iitb.ac.in}
\affiliation{Department of Physics, Indian Institute of Technology Bombay, Mumbai 400076, India}

\author{Aftab Alam}
\email{aftab@phy.iitb.ac.in}
\affiliation{Department of Physics, Indian Institute of Technology Bombay, Mumbai 400076, India}


\begin{abstract}

We report the theoretical prediction of a new class of spintronic materials, namely bipolar magnetic semiconductor (BMS), which is also supported by our experimental data. BMS acquires a unique band structure with unequal band gaps for spin up and down channels, and thus are useful for tunable spin transport based applications such as spin filters. The valence band (VB) and conduction band (CB) in BMS approach the Fermi level through opposite spin channels, and hence facilitate to achieve reversible spin polarization which are controlable via applied gate voltage. We report the quaternary Heusler alloy VNbRuAl to exactly possess the band structure of BMS. The alloy  is found to crystallize in LiMgPdSn prototype structure (space group $F\bar{4}3m$) with B$2$ disorder and lattice parameter 6.15 \AA . The resistivity and Hall measurements show a two channel semiconducting behavior and a quasi linear dependance of negative magneto resistance (MR) indicating the possible semiconducting nature. Interestingly, VNbRuAl also shows a fully compensated ferrimagnetic (FCF) behavior with vanishing net magnetization (m$_s$$\sim$ $10^{-3}$ $\mu_B/f.u.$) and significantly high ordering temperature ($> 900$ K). Unlike conventional FCF, vanishing moment in this case appears to be the result of a combination of long range antiferromagnetic (AFM) ordering and the inherent B2 disorder of the crystal. This study opens up the possibility of finding a class of materials for AFM spintronics, with great significance both from fundamental and applied fronts.

\end{abstract}


\date{\today}

\maketitle
\begin{figure}[b]
 \centering
 \includegraphics[width=0.99\linewidth]{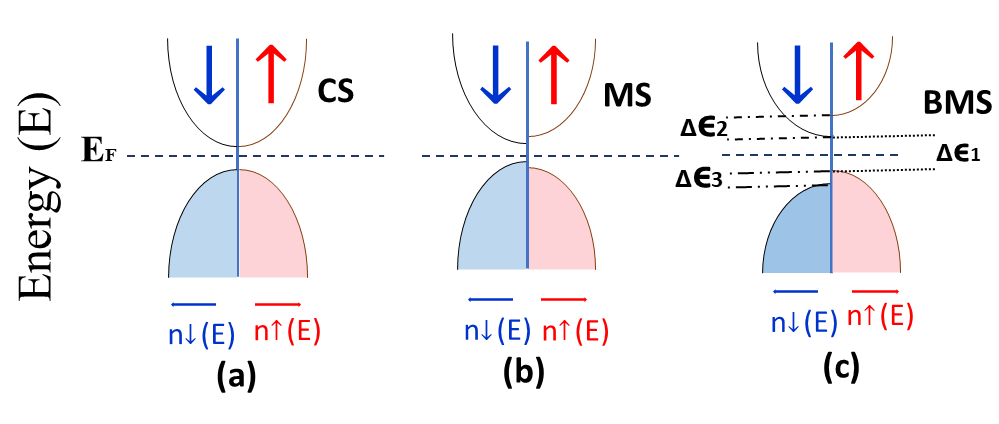}
 \caption{Schematic of density of states $n(E)$, for a typical (a) conventional semiconductor (CS) (b) magnetic semiconductor (MS) and (c) bipolar magnetic semiconductor (BMS).}
 \label{fig:schematic-SGS-SM}
 \end{figure}

{\it\bf{Introduction:}}
Heusler alloys are exhilarating due to their exotic physical properties with potential applications in areas such as spintronics, topological quantum matter, spin filters etc. They are known to have excellent stability, high Curie temperature ($T_{\mathrm{C}}$), high spin polarization\cite{rani2019spin} and compatibility to grow thin films, etc.\cite{venkateswara2019coexistence} Spintronic technology has several advantages over the conventional electronics\cite{Wolf1488,vzutic2004spintronics,felser2007spintronics,graf2011simple,prinz1998magnetoelectronics} such as low power consumption, high storage density, high processing speed, etc. A wide variety of materials have been reported in the literature  to achieve this goal. It started with a class of materials called ``half metallic ferromagnets'' (HMF),\cite{venkateswara2020half,venkateswara2015electronic} in which one of the spin band is metallic while the other is semiconducting or insulating. Later, spin gapless semiconductors (SGS)\cite{bainsla2015spin,xu2013new,bainsla2015origin} were discovered, which gained prominence over HMF due to their unique band properties. Magnetic semiconductors (MS)\cite{stephen2019structural} and spin semi-metals \cite{venkateswara2019coexistence} constitute the other unique classes. A number of these class of compounds have been reported by our group in the past.\cite{venkateswara2019coexistence,
venkateswara2020half,venkateswara2015electronic,bainsla2015spin,bainsla2015origin}  In this letter, we report another interesting class of materials, namely bipolar magnetic semiconductors (BMS). BMS bridge the gap between conventional semiconductors(CS) and magnetic semiconductors(MS). CS has equal band gaps in both spin channels whereas MS possesses a band structure with unequal band gaps for spin up and down channels. BMS acquire a similar band structure as MS with a key difference that the VB and CB shift towards the Fermi level (E$_F$) in the opposite direction with respect to spin. This unique electronic feature enables BMS to achieve reversible spin polarization which can be controlled via an applied gate voltage (V$_g$). Schematic density of states (DoS) for CS, MS and BMS are shown in Fig. \ref{fig:schematic-SGS-SM}. BMS can be characterized by three important parameters $\Delta {\epsilon_1}$, $\Delta{\epsilon_2}$, and $\Delta{\epsilon_3}$. $\Delta{\epsilon_1}$ is the energy gap between VB and CB edges of spin up and down channels respectively, while $\Delta{\epsilon_2}$($\Delta{\epsilon_3}$) are the gaps between CB (VB) edges  of the two spins, as depicted in Fig. \ref{fig:schematic-SGS-SM}(c).
Also, $\Delta{\epsilon_1}$ + $\Delta{\epsilon_2}$ (= E$_g$$\uparrow$) and $\Delta{\epsilon_1}$ + $\Delta{\epsilon_3}$ (= E$_g$$\downarrow$) quantifies the band gaps for the spin up and down channels respectively.
One of the potential technologies where BMS can be extremely useful is spin filters in which it is possible to tune E$_F$ into either spin up VB or spin down CB by varying V$_g$, and hence, achieve a tunable spin polarization.  \\

Although there exists a few theoretical proposals for BMS in the literature,\cite{li2012bipolar,zhang2015electrical,du2017novel} no experimental confirmation is reported so far. The purpose of this letter is to first establish the BMS property in a quaternary Heusler alloy, VNbRuAl, using first principles calculations. Next, we demonstrate some of the  indirect experimental evidences to confirm the necessary features of BMS in this alloy. In addition to BMS, this compound is experimentally confirmed to be a fully compensated ferrimagnet (FCF) with vanishing magnetization. FCF possesses several advantages over other conventional spintronic materials, namely (i) no external stray fields resulting into low energy losses, (ii) better spin sensitivity, which allows them not to disturb spin characters and make them ideal for spin polarized scanning tunneling microscope tips, and (iii) low shape anisotropy, useful in the applications of spin injection. Interestingly, VNbRuAl is a classic system which violates the well-known Slater-Pauling (SP) rule,\cite{ozdougan2013slater,zheng2012band,gao2015first} according to which this system should have a finite magnetization of 3 $\mu_B$.  A microscopic origin behind such discrepancy is explained by our ab-initio calculations. We attribute such a vanishing of magnetization to either of the two factors; (i) long range magnetic ordering with specific spin orientations of different magnetic atoms or (ii) B2 disorder, as confirmed from our powder x-ray diffraction (XRD) data. Details of the sample preparation, experimental tools and computational methods, used to study the present system, are given in the supplemental material (SM).\cite{supplement}

\begin{center}
 \bf{THEORETICAL RESULTS}
\end{center}
 
 VNbRuAl crystallizes in LiMgPdSn prototype structure (space group $F\bar{4}3m$) with experimentally measured lattice parameter of 6.15 \AA. The structure can be seen as 4 interpenetrating fcc sublattices with Wyckoff positions 4$a$, 4$b$, 4$c$ and 4$d$. In general, there exists three energetically non-degenerate lattice configurations for  XX$'$YZ alloy (fixing Z-atom at 4$a$-site). They are
(I) X at 4$c$, X$'$ at 4$d$ and Y at 4$b$ site,
(II) X at 4$b$ , X$'$ at 4d and Y at 4c site, and
(III)X at 4$c$ , X$'$ at 4$b$ and Y at 4$d$ site.

\begin{table}[t]
 	\centering
 	\caption{For VNbRuAl, relaxed lattice parameters ($a_0$), total and atom-projected magnetic moments (in $ \mu_B$) and relative energy ($\Delta E$) of the three configurations with reference to type-III configuration within HSE06 functional.}
 	\begin{tabular}{l c c c c c c}
 		\hline \hline
 		Type&  $a_0$ (\AA) $ \ $  &  $m^{\mathrm{V}}$ & $\ $ $m^{\mathrm{Nb}}$ $\ $  &  $m^{\mathrm{Ru}}$  & $\ $ $m^{\mathrm{Total}}$ $ \ $ & $\Delta E$(eV/f.u.) \\ \hline
 		I   &  6.26   & 2.53 	& 	 0.38   	& 	 -0.28 	& 2.6		&  1.24   \\
 		II  &  6.19   & 0.00		& 0.00 	& 	0.00 	&  0.00		& 0.7    \\
 		III    &  6.24  &  2.72 	&     -0.10  	&	 0.18 	& 2.80 		&  0   \\
 		\hline \hline
 	\end{tabular}
 	\label{tab:theory-VNRA}
 \end{table}

Using first principles calculations, all three non-degenerate configurations (I, II and III) were fully relaxed with various spin orientations (ferromagnetic, antiferromagnetic and ferrimagnetic). Type III turns out to be energetically the most stable one for VNbRuAl.
These calculations were first performed using Perdew, Burke, and Ernzerhof (PBE) exchange-correlation functional,\cite{perdew1996generalized} details of which are given in SM.\cite{supplement} Keeping in mind the shortcomings of PBE, we have repeated all the above calculations using the hybrid HSE06 functional,\cite{krukau2006influence}  which is known to provide more accurate electronic structure, though computationally more expensive. Table \ref{tab:theory-VNRA} shows the relaxed lattice parameters, total and atom-projected moments and the relative energy differences among the three structural configurations.

\begin{figure}[t]
 	\centering
 	\includegraphics[width= 1.1\linewidth]{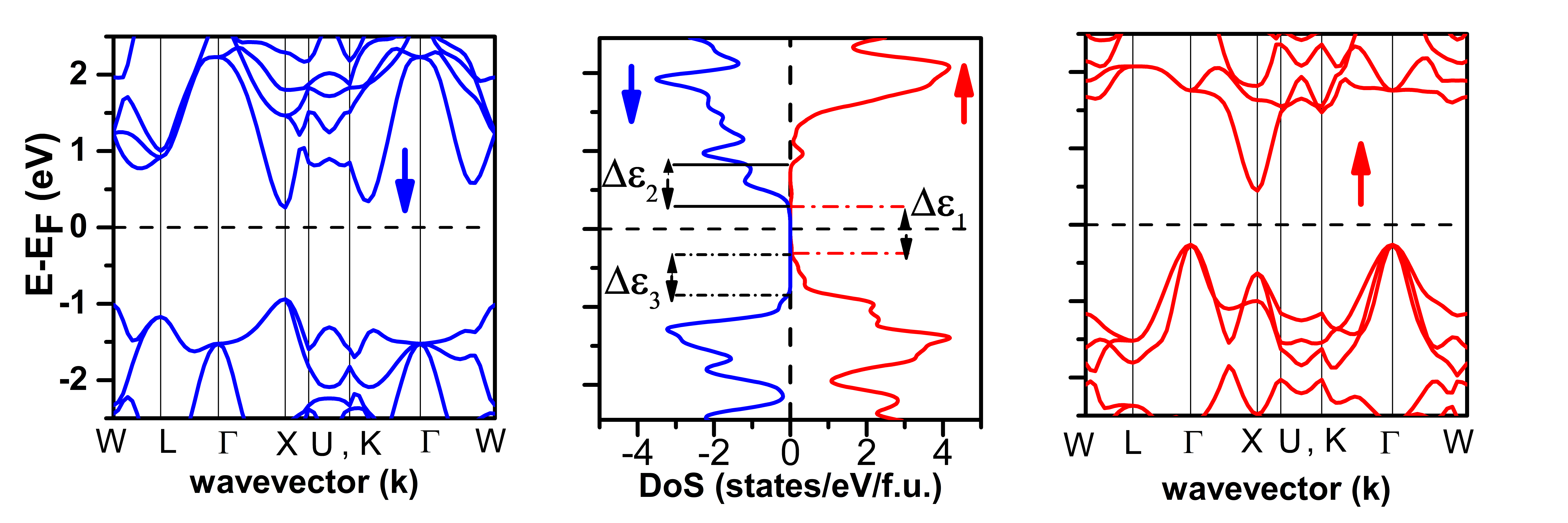}
 	\caption{For VNbRuAl (type III configuration) spin polarized band structure and density of states at relaxed lattice parameter a$_0$. DoS plot clearly elucidates the concept of BMS highlighting the energy parameters $\Delta {\epsilon_1}$, $\Delta{\epsilon_2}$, and $\Delta{\epsilon_3}$.}
 	\label{fig:DoS-VNRA}
 \end{figure}

The theoretically optimized lattice parameter, $a_0=6.24$ \AA, matches fairly well with experiment, and the net moment ($m^{\mathrm{Total}}$=2.8 $ \mu_B$) nearly follows the SP rule (3 $\mu_B$). Although type III configuration turns out to be most stable from both PBE and HSE calculations, the moments and the total energies for the three configurations are clearly different.
Figure \ref{fig:DoS-VNRA} shows the spin polarized band structure and DoS for type III configuration using HSE functional. Unlike PBE results, HSE calculation shows the BMS behaviour for VNbRuAl. The DoS plot also highlights the 3 energy parameters  $\Delta {\epsilon_1}$, $\Delta{\epsilon_2}$, and $\Delta{\epsilon_3}$, characterising the BMS feature, as listed in Table \ref{tab:theory1-VNRA}. From application point of view, it is desirable to achieve small value for  $\Delta {\epsilon_1}$ and relatively large values for E$_g$$\uparrow$ and E$_g$$\downarrow$, which is indeed the case for VNbRuAl.

\begin{table}[b]
 	\centering
 	\caption{For type III configuration of VNbRuAl, relaxed lattice parameter $a_0$ (in \AA), energy parameters (all in eV)  $\Delta {\epsilon_1}$, $\Delta{\epsilon_2}$, $\Delta{\epsilon_3}$, and band gaps for majority (E$_g$$\uparrow$=$\Delta {\epsilon_1}$ + $\Delta {\epsilon_2}$) and minority (E$_g$$\downarrow$=$\Delta {\epsilon_1}$ + $\Delta {\epsilon_3}$) spins using the HSE06 functional.}
 	\begin{tabular}{l c c c  c c c}
 		\hline \hline
 		Type $ \ \ \ \ $ &  $a_0$ $ \ \ \ \ $  &  $\Delta {\epsilon_1}$ $ \ \  \ $ &  $\Delta {\epsilon_2}$ $ \ \ \ \ $ &  $\Delta {\epsilon_3}$ $ \ \ \ \ $& E$_g$$\downarrow$  $ \ \ \ \ $ & E$_g$$\uparrow$   \\ \hline
 		III    $ \ \ \ \ $&  6.24  $ \ \ \ \ $&  0.65 	$ \ \ \ \ $&     0.52  	$ \ \ \ \ $&	0.47 $ \ \ \ \ $& 1.12 $ \ \ \ \ $&  1.17 \\
 		\hline \hline
 	\end{tabular}
 	\label{tab:theory1-VNRA}
 \end{table}

\begin{center}
 \bf{EXPERIMENTAL RESULTS}
\end{center}
 {\it\bf {Crystal Structure: }} Figure \ref{fig:xrd-VNRA} shows the room temperature XRD pattern for VNbRuAl along with the Rietveld refinement for configuration III with 50\% disorder between V-Ru and Nb-Al atoms. Inset (ii) shows a zoomed view of the above fitting near (111) and (200) peaks. Inset (i) shows the refined data with pure configuration III (with no disorder), which clearly does not fit well. We have also fitted other structural configurations (I and II) in pure form as well as their disorder variants (octahedral or tetrahedral sites individually and both together), however, the best fit with the lowest $\chi^2$ value of 2.9 was obtained for configuration III with 50\% disorder.  Details of XRD analyses including these refinements are shown in SM.\cite{supplement} These results confirm that VNbRuAl crystallizes in configuration III with B2 disorder, and experimental lattice constant 6.15 \AA.

 {\it\bf {Magnetic properties:}}
According to the SP rule, \cite{ozdougan2013slater,zheng2012band,gao2015first} the saturation magnetization in the present case is given by M$_S$ = (N$_v$ - 24)~$\mu_B/f.u.$, where $N_v$ is the total number of valence electrons of the Heusler alloy. This yields the value of M$_S$ = 3 $\mu_B/f.u.$ for VNbRuAl.

\begin{figure}[t]
\centering
\includegraphics[width=0.9\linewidth]{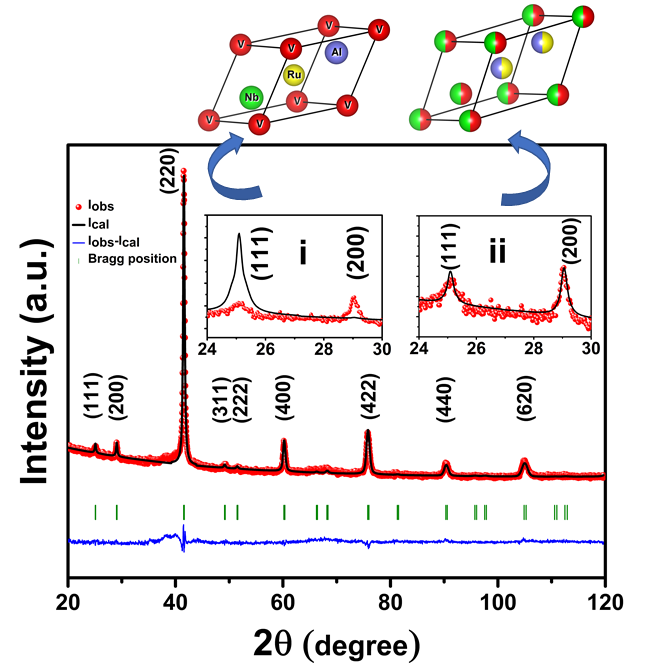}
\caption{Room temperature powder XRD pattern of VNbRuAl along with the Rietveld refined data for configuration III with B$2$ disorder between V-Ru and Nb-Al sites. Inset (ii) shows a zoomed-in view near (111) and (200) peaks of XRD pattern for the above case. Inset (i) shows the zoomed-in view when pure configuration III ( with no disorder) is fitted.}
\label{fig:xrd-VNRA}
\end{figure}

\begin{figure}[t]
\centering
\includegraphics[width=1.05\linewidth]{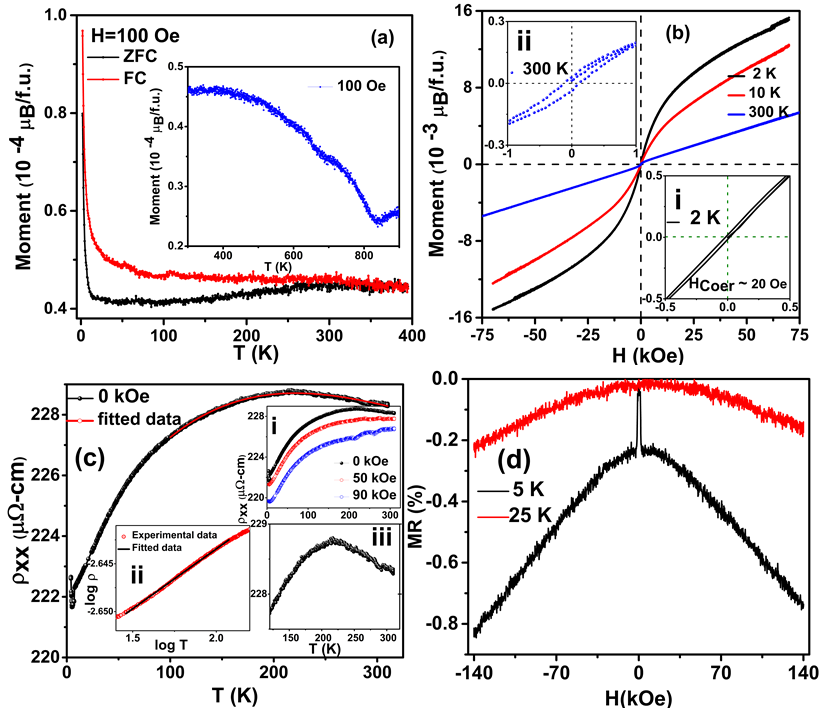}
\caption{For VNbRuAl (a) M vs. T in zero-field cooled (ZFC) and field cooled (FC) modes in H=100 Oe. The inset shows the high-temperature regime. (b) M vs. H at 2, 10 and 300 K . The insets (i) and (ii) display zoomed-in view at 2 K and 300 K respectively. (c) Resistivity ($\rho_{xx}$) vs. T, along with a two carrier model fit in 0 kOe. The insets (i) shows resistivity ($\rho_{xx}$) vs. T in three different fields (ii) shows plot of $\log \rho $ vs. $\log T$ fitted in the temperature range between $30~K - 110~K$ and (iii) shows a zoomed-in view in the high (T $>$ 100 K) temperature regime of the resistivity. (d) The magnetoresistance (MR) vs. H at 5 K and 25 K.}
\label{fig:VNRA-MH-MT}
\end{figure}

Figures \ref{fig:VNRA-MH-MT}(a) and \ref{fig:VNRA-MH-MT}(b) show M vs. T and M vs. H curves respectively for VNbRuAl. Interestingly, the net moment turns out to be negligibly small ($\sim$$10^{-3}$ $\mu_B$), which is in complete contrast to the prediction of SP rule. Inset of Fig. \ref{fig:VNRA-MH-MT}(a) clearly suggests large ordering temperature ($>$ 900 K), much larger than other reported FCF materials.\cite{PhysRevApplied.7.064036} The M-H curve at 2K shows a small, but non-zero hysteresis, with a small coercivity of 20 Oe (see inset (i) of Fig. \ref{fig:VNRA-MH-MT}(b)). In particular, the M-H curve is non-saturating up to 90 kOe, indicating the compensated ferrimagnetic like behavior.\cite{stephen2016synthesis} There can be various plausible reasons for such vanishing moment, e.g. (i) compensated ferrimagnetism (ii) long range antiferromagnetism (iii) the B2 disorder as predicted by XRD.

\begin{center}
{\bf {TRANSPORT PROPERTIES}}
\end{center}

{\it\bf Resistivity:}
Figure \ref{fig:VNRA-MH-MT}(c) shows the temperature dependence of the longitudinal resistivity ($\rho_{xx}$) at different fields (inset (i) of Figure \ref{fig:VNRA-MH-MT}(c)). The resistivity increases with T from the low temperature region and goes through a maximum $\sim$ 220 K before falling as T increases further towards 300 K. The temperature variation of $\rho_{xx}$ is quite similar to few other Heusler systems \cite{jamer2013magnetic,nishino1997semiconductorlike} with semiconducting behavior. On the onset, it appears that low T-region of the $\rho_{xx}$ is metallic due to phonon scattering, and beyond 200 K, it shows semiconducting like behavior. The peak at $\sim$ 220 K (see inset (iii)) could be due to the competition between positive temperature coefficient of resistivity ($ d\rho / dT > 0 $) arising from metallic character followed by a region of negative temperature coefficient of resistivity ($d\rho / dT < 0 $) at higher T due to semiconducting behaviour. In many semiconducting materials with metallic like behaviour at low T\cite{jamer2013magnetic,nishino1997semiconductorlike}, the low T resistivity is found to vary as  T$^\alpha$ with $\alpha$-values lying between 2 and 3. In order to check this, we have plotted $\log \rho_{xx} $ vs. $\log T$ and fitted a straight line (see inset (ii)). Such curve fits well only in the T-range of $30~K - 110~K$ and the value of $\alpha$ turns out to be 0.012. This value of $\alpha$ is surprisingly very small, which indicates that the increase in $\rho_{xx}$  with T is negligibly small, and hence the metallic character is extremely weak.
It is well known that, in semiconductors, valence electrons can surmount the gap by absorbing a small amount of energy, and more electron-hole pairs are created as T is increased. These thermally activated carriers are possibly responsible for the semiconducting behavior of VNbRuAl and is the dominating mechanism at higher T. The conductivity ($\sigma(T)$) data is fitted in the T-range of 120-320 K with the modified two-carrier model,\cite{kittel2007introduction,jamer2017compensated}
\begin{equation}
\sigma(T) = e (n_e \mu_e + n_h \mu_h)
\label{eq:tbm}
\end{equation}
where, $n_i=n_{i0}\ e^{-\Delta E_i/k_\mathrm{B}T}$ ($i=e, h$) are the carrier concentration of electrons, holes with mobilities $\mu_i$ and pseudo-gaps $\Delta E_i$. $\mu_i$ can be written as $\mu_i = (a{_i}T + b_i)^{-1} = \mu_{i0}/({a'{_i}}T + 1) $. Here $a$ corresponds to carrier-phonon scattering while $b$ arises from the mobility due to defects at T=0 K.
Eq. (\ref{eq:tbm}) then take the form,
\begin{equation}
\sigma(T) = [A_e(T) \ e^{-\Delta E_e/k_\mathrm{B}T} + A_h(T) \ e^{-\Delta E_h/k_\mathrm{B}T}].
\label{eq:tbm-final}
\end{equation}
where, $ A_i(T) = en_{i0}\mu_{i0}/({a'{_i}}T + 1) $. Equation(\ref{eq:tbm-final}) is used to fit the zero-field  $\rho$-data between 100 to 320 K (see Fig. \ref{fig:VNRA-MH-MT}(c)), and the energy gaps ($\Delta E_i$) turn out to be 64.3 meV and 0.23 meV. The very fact that two carrier model describes $\rho$-data over such a wide temperature range, indicates that semiconducting behaviour is dominated from 100~K itself. Well below 100 K, phonon scattering dominates the electrical transport in VNbRuAl. Interestingly, the zero field $\rho$-peak at $\sim$ 220 K gets suppressed at higher fields, giving rise to a much reduced variation with T.\cite{SHAPIRA1971272} The order of magnitude of $\rho$ in VNbRuAl is quite similar to other semiconducting materials reported in the literature.\cite{venkateswara2018competing}

\begin{table}[t]
	\centering
	\caption{For VNbRuAl, relative energy (in eV/f.u.) of the three disordered configurations with reference to type III configuration ($\Delta$E$_{tot}$), average spin up and down magnetic moments (in $\mu_B$) at V, Nb and Ru atoms.  }
	\begin{tabular}{l c c c c c c c}
		\hline \hline
		Type &  $\Delta$E$_{tot}$   &\ \ $m^{V}_{\uparrow}$ \ \ & \ \ $m^{V}_{\downarrow}$\ \  & $\ \ m^{Nb}_{\uparrow}$\  \ & \ \ $m^{Nb}_{\downarrow}$\ \ & $\ \ m^{Ru}_{\uparrow}$\ \ & $m^{Ru}_{\downarrow}$   \\ \hline
		I   &  0.434   & 0.00 	& 0.00 & 0.00 	& 0.00 & 0.00 	& 0.00    \\
		II   &  0.136   & 0.00 	& 0.00 & 0.00 	& 0.00 & 0.00 	& 0.00    \\
		III   &  0.0   & +0.73 	& -0.62 & +0.01 	& -0.02 & +0.04 	& -0.01    \\
		\hline \hline
	\end{tabular}
	\label{tab:theory2-VNRA}
\end{table} 

{\it\bf Magnetoresistance:}
Figure \ref{fig:VNRA-MH-MT}(d) shows the field dependence of magnetoresistance, MR(H) = $ \left[ \rho(H) - \rho(0)\right] /\rho(0)$, at two different temperatures for VNbRuAl. The MR decreases with increasing magnetic field and a sharp peak like feature is observed in 5K data. This peak lies within $\pm$2.5~kOe. Sensitivity of this peak with  magnetic field suggests its connection with magnetism of the sample. There is a similar sharp rise in $\rho_{xx}$  below 6~K in zero-field resistivity, which gets suppressed at higher H, and sharp rise of both ZFC and FC magnetization (see Fig. \ref{fig:VNRA-MH-MT}(a)). These observations seem to suggest some kind of spin reorientation below 6~K. Therefore, at 5~K the spins might be misaligned to give higher resistivity but when field aligns them, resistivity drops sharply. With increasing T, the variation of MR with H gets reduced significantly. In Figure \ref{fig:VNRA-MH-MT}(d), MR shows a quasi linear dependence on H, without any sign of saturation till 140 kOe, even at 5~K. The slope of this linear regime decreases with increasing T. Such behaviour reflects a gapped semiconductor,\cite{xu1997large} as reported in various systems by Abrikosov.\cite{PhysRevB.58.2788}

{\it\bf Hall Effect:}
The Hall resistivity $\rho_{xy}$ vs. field, at different T, shows a linear dependence (see Figure S4 in SM\cite{supplement}). The calculated carrier concentration $(n)$ at 5 K is  3.5 x $10^{19}$ cm$^{-3}$, which is comparable to that of other Heusler alloys having semiconducting nature.\cite{rani2019spin} The carrier concentration increases marginally (from 3.5$\times$10$^{19}$ to 4.7$\times$10$^{19}$) with increasing T, again indicating the robust semiconducting like behavior of VNbRuAl. The positive slope of the ordinary Hall coefficient suggests that holes are the majority charge carriers.

\begin{figure}[t]
	\centering
	\includegraphics[width= 1.0\linewidth]{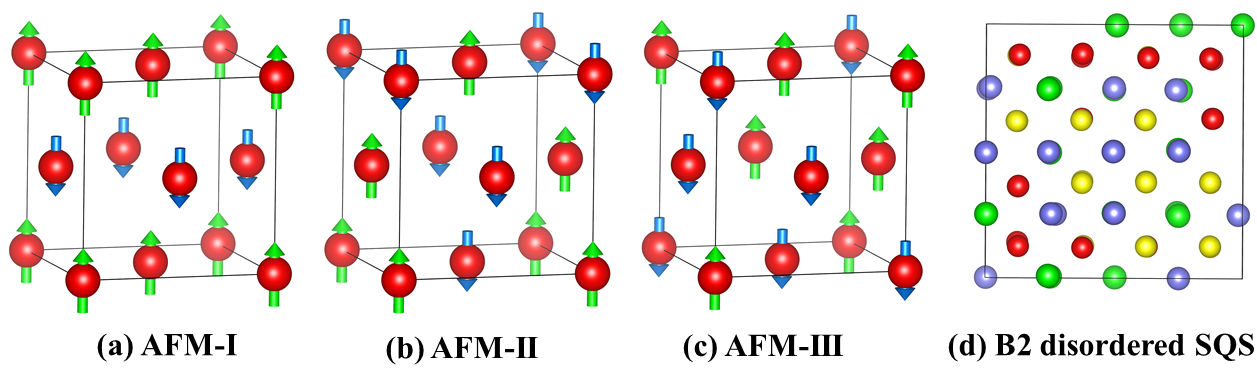}
	\caption{Three different antiferromagnetic states, (a) AFM-I, (b)AFM-II and (c) AFM-III, displaying only V-atoms in the unit cell. (d) 64-atom SQS structure with B2 disorder in type III configuration of VNbRuAl. }
	\label{fig:VNPicture2}
\end{figure}

\begin{center}
\bf{EXPLAINING FCF BEHAVIOR}
\end{center}

Magnetisation measurements predict vanishing moments for VNbRuAl, which is in contrast to the theoretical prediction (2.8 $ \mu_B$) as well as SP rule (3 $\mu_B$). Such vanishing magnetization may arise due to various reasons such as (1) a unique AFM ordering and/or (2) disorder.
To better understand the experimental observation, we first simulated a few distinct AFM configurations involving V-atoms, as shown in Fig.\ref {fig:VNPicture2}. Structural optimization confirms FM phase to be energetically the most stable, as shown in Figure S1(a) of SM.\cite{supplement} Interestingly, AFM-III phase lies only $\sim$10 meV/atom above FM phase. To cross check the effect of varying exchange correlation functionals, we have optimized all these phases within local density approximation (LDA) as well (see Fig. S1(b) of SM\cite{supplement}). This predicts FM and AFM-III to be energetically degenerate, and hence AFM-III phase to be equally probable in the real sample giving rise to a vanishing moment.

We further simulated the effect of B2 disorder (as confirmed by our XRD data, Fig. \ref{fig:xrd-VNRA}) on magnetic properties of VNbRuAl. To simulate B2 disorder, we generated special quasi random structures (SQS),\cite{zunger1990special} which are known to mimic the random correlations. Figure \ref {fig:VNPicture2}(d) shows a 64-atom SQS cell for B2 disorder in configuration III.  We have actually generated B2 disordered structures for all three  configurations (I, II and III). Table \ref{tab:theory2-VNRA} shows the relative energies between the three disordered structures ($\Delta$E$_{tot}$) with reference to type III disorder, and the average spin up and down moments on V, Nb and Ru atoms. From Table \ref{tab:theory2-VNRA}, one can see that B2 disorder in the type III configuration is energetically most favorable, and hence highly likely to occur in the synthesized sample, as confirmed by our experiment.  
Interestingly, our simulation reveals a drastically reduced net magnetization in the type-III B2 disordered configuration (net magnetization is the sum of spin up and down magnetization at all atoms), which explains the observed FCF behavior. Most importantly, unlike AFM case, the simulated local moments in disordered phase confirms the ferrimagnetic ordering on V atoms having different electronic structure for spin up and down bands. This is one of the prerequisites for BMS behaviour.

\begin{center}
\bf{SUMMARY AND CONCLUSION}
\end{center}
In this letter, we report a new class of spintronic materials
which showcases the co-existence of two composite quantum phenomena, namely bipolar magnetic semiconductor (BMS) and fully compensated ferrimagnetism. Using a combined theoretical and experimental study, we confirm VNbRuAl to be the first system to exhibit the above unique property. VNbRuAl crystallizes in cubic
Heusler structure with B2 disorder. Ab-intio calculations predict VNbRuAl as a bipolar magnetic semiconductor with ferrimagnetic ordering. Transport measurements indicate two parallel channels in both resistivity and magnetoresistance, confirming the semi-
conducting character. Magnetization measurements show a vanishing net moment, confirming the FCF behavior
of VNbRuAl. This is however an unconventional FCF, which arises due to a combination of long range antiferromagnetic ordering and the inherent B2 disorder of the crystal, as explained by our first principles simulation. Our results suggest that such a co-existence of different interesting properties (BMS, FCF and high T$_C$) in a single compound opens up new opportunities for spintronics applications e.g. room temperature spin filters.


{\it \bf Acknowledgments:} JN and JK acknowledge the financial support, in the form of TAship, provided by IIT Bombay. AA acknowledges DST-SERB (Grant No. MTR/2019/ 000544) for funding to support this research.



\bibliographystyle{apsrev4-1}
\bibliography{references}

\pagebreak
\clearpage
\renewcommand{\bibnumfmt}[1]{[S#1]}
\renewcommand{\citenumfont}[1]{S#1}

\begin{center}
\textbf{\large Supplementary Material for ``Bipolar Magnetic Semiconducting Behavior in VNbRuAl: A New Spintronic Material for Spin Filters''}
\end{center}

Here, we present the details of the sample preparation, experimental tools and computational methods used to study the Heusler alloy VNbRuAl. In the theoretical results section, we presented the electronic structure of three different structural configurations (type I, II and III) within PBE functional. We also presented the energetics of various antiferromagnetic phases for type III configuration.  We have demonstrated the details of the XRD refinement and the Hall resistivity $\rho_{xy}$ vs. field (H) data for VNbRuAl in the experimental results section.

 \section{Experimental Details}
  Polycrystalline samples of VNbRuAl were synthesized using arc melting in a high purity argon environment of the constituents in stoichiometric proportion and with a purity of 99.99\%. To obtain single phase, after melting, the samples were annealed for 2 weeks at 800$^{\circ}$C in sealed quartz tubes, followed by furnace cooling. Room temperature X-ray diffraction (XRD) patterns were taken using Cu-K$\alpha$ radiation with the help of Panalytical X-pert diffractometer. Crystal structure analysis was done using FullProf Suite software.\cite{rodriguez1993recent} Magnetization measurements at various temperatures were performed using a vibrating sample magnetometer (VSM) attached to the physical property measurement system (PPMS) (Quantum design) in fields up to 70 kOe. High-temperature magnetization measurements (300-900 K) were carried out using a superconducting quantum interference device magnetometer (magnetic property measurement system, Quantum Design). Temperature and field dependent resistivity measurements were carried out using the PPMS (DynaCool, Quantum Design) with the electrical transport option (ETO) in the traditional four probe method, applying a 10 mA current at 18 Hz frequency. In order to understand the type of charge carriers and the variation of their density with temperature, Hall measurements were performed using the PPMS with the van der Pauw method again using 10 mA of current at 18 Hz frequency.
 
 \section{Computational details}

 {\it Ab-initio} calculations were performed using the spin resolved density functional theory (DFT),\cite{hohenberg1964inhomogeneous} as implemented within Vienna ab initio simulation package (VASP) \cite{kresse1996efficient,kresse1996efficiency,kresse1993ab} with a projected augmented-wave (PAW) basis.\cite{kresse1999ultrasoft} Pseudopotential formalism with Perdew, Burke, and Ernzerhof (PBE) exchange-correlation functional\cite{perdew1996generalized} was used for primary electronic structure calculations. To perform the Brillouin zone integration within the tetrahedron method, a $24\times 24\times 24$ $\Gamma$-centered k-mesh was used. A plane wave energy cut-off of 500 eV was used for all the calculations. All the structures are fully relaxed with total energies (forces) converged to values less than 10$^{-6}$ eV (0.01 eV/\AA). PBE functional is known to underestimate the band gap of semiconducting materials. And, in the present case, because we are dealing with magnetic semiconductors, it is essential to go beyond PBE calculation. To make a more accurate prediction, we employed the Heyd-Scuseria-Ernzerhof (HSE06) \cite{krukau2006influence} functional. A 16 atom conventional cell and its symmetric $2\times 2\times 2$ supercell have been used while simulating ferromagnetic (FM) and different antiferromagnetic (AFM) configurations. To incorporate the B2-disorder, we have generated a 64 atom special quasirandom structure (SQS).\cite{zunger1990special} SQS is an ordered structure, known to mimic the random correlation accurately, for disordered compounds. Alloy Theoretic Automated Toolkit (ATAT)\cite{van2013efficient} was used to generate the SQS structures. The generated SQS structures perfectly mimic the random pair correlation functions up to third nearest neighbors.
 
 \section{Theoretical Results}
\begin{figure}[t]
\renewcommand{\thefigure}{S\arabic{figure}}
 \centering
 \includegraphics[width=1.0\linewidth]{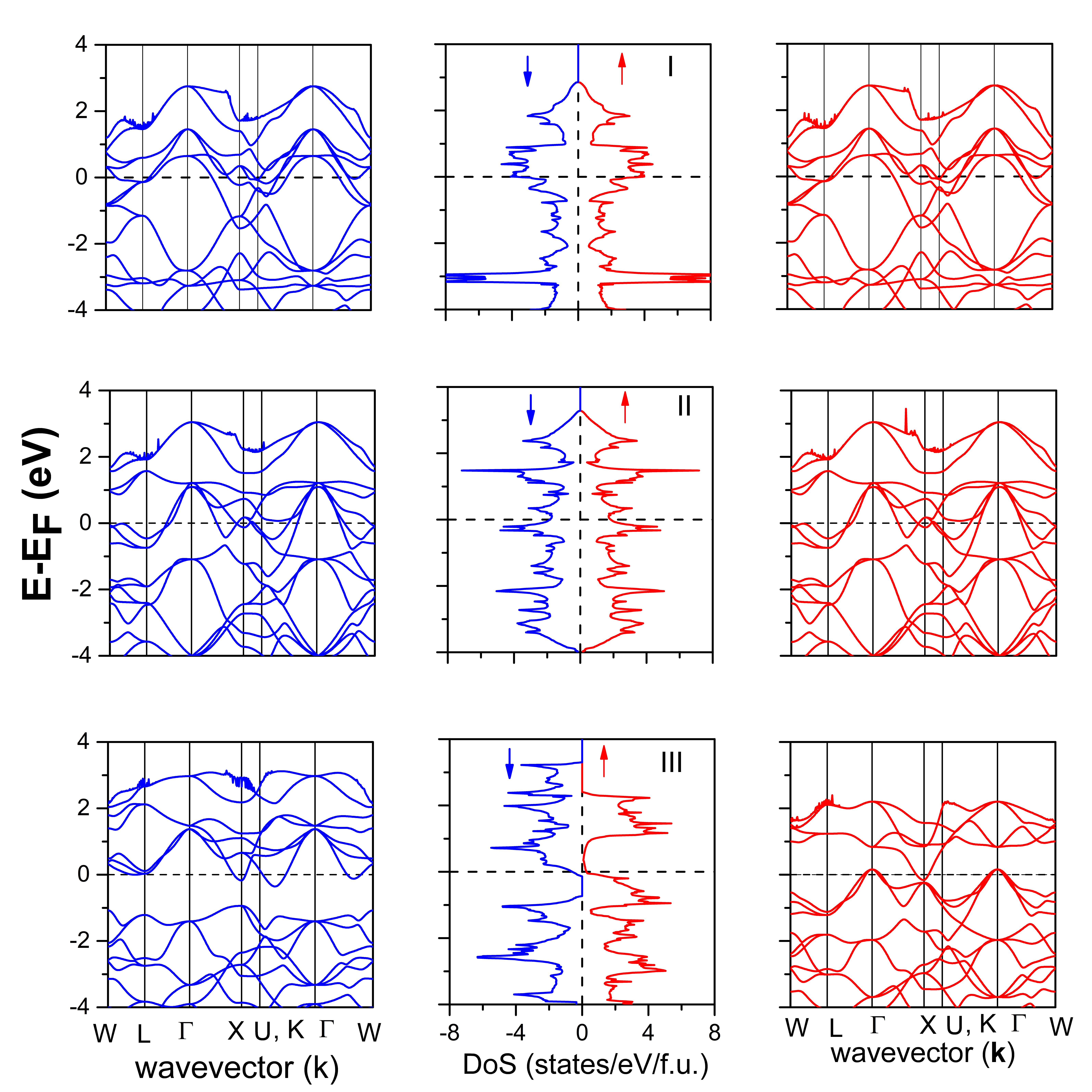}
 \caption{Spin polarized band structure and DoS as obtained using the PBE functional (within the GGA approximation) for the three configurations I, II and III of VNbRuAl at their respective relaxed lattice parameters a$_0$. Left panel corresponds to the spin down band while the right panel for spin up band structure.}
 \label{fig:VNRA-pbe1}
 \end{figure}

Various magnetic states were examined within ab-initio framework by considering different initial spin (nonmagnetic, ferro- and ferri-magnetic) arrangements for all the three configurations, I, II and III. Table \ref{tab:theory-VNRA} shows the relaxed lattice parameters, total and atom projected moments and relative energy difference between the three configurations  using the PBE functional (within the GGA approximation). One can notice that, Type-III configuration  with ferromagnetic ordering is energetically the most stable configuration.  Spin resolved band structure and density of states (DoS) for all the three configurations are shown in the Fig. \ref{fig:VNRA-pbe1}. Type I and II configurations show metallic behavior while type III show nearly metallic state.  Theoretically relaxed lattice parameter ($a_0=6.24$ \AA) for the most stable configuration matches well with the experimentally measured value. The calculated net magnetization is $2.65\ \mu_B$/f.u.

\begin{figure}[t]
 \renewcommand{\thefigure}{S\arabic{figure}}
 \centering
 \includegraphics[width=1.0\linewidth]{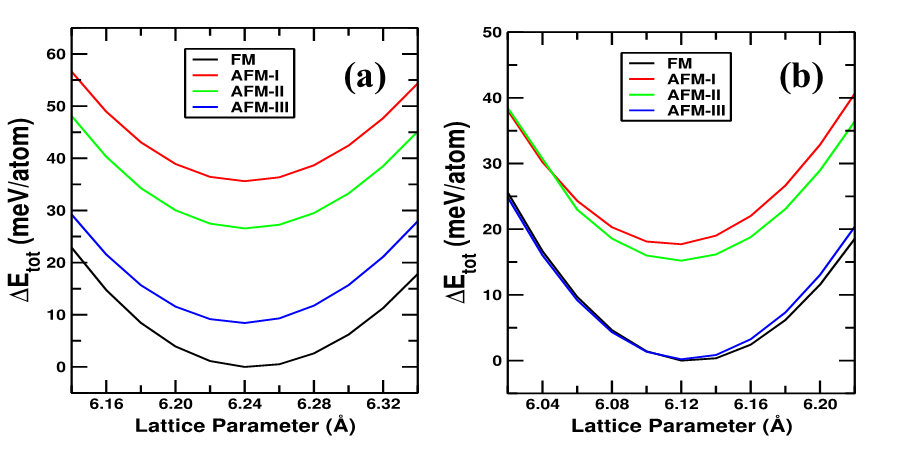}
 \caption{Simulated relative energy difference between ferromagnetic and different antiferromagnetic states of VNbRuAl using (a) PBE and (b) LDA exchange correlation functionals.}
 \label{fig:S1}
 \end{figure}

Figure \ref{fig:S1} shows the energy minimization curves for different magnetic states (ferromagnetic (FM) and three different antiferromagnetic (AFM) states), a schematic diagram for which are shown in Fig. 5 of the main manuscript. The left (right) panel shows the results  using PBE (LDA) exchange correlation functional. Clearly, FM is energetically most stable configuration within the PBE functional. AFM-III phase is only 10 meV above the FM phase. In contrast, within LDA functional, FM and AFM-III are energetically degenerate indicating that AFM-III is equally probable to be present in the real sample of VNbRuAl.

 \begin{table}[t]
 \renewcommand{\thetable}{S\arabic{table}}
 	\centering
 	\caption{For VNbRuAl, relaxed lattice parameters ($a_0$), total and atom-projected magnetic moments (in $ \mu_B$) and relative energy ($\Delta E$) of the three configurations with reference to type-III configuration within PBE functional.}
 	\begin{tabular}{l c c c c c c}
 		\hline \hline
 		Type&  $a_0$ (\AA) $ \ $  &  $m^{\mathrm{V}}$ & $\ $ $m^{\mathrm{Nb}}$ $\ $  &  $m^{\mathrm{Ru}}$  & $\ $ $m^{\mathrm{Total}}$ $ \ $ & $\Delta E$(eV/f.u.) \\ \hline 
 		I   &  6.23   & 0.02 	& 	 0.00   	& 	 0.00 	& 0.02		&  1.27   \\
 		II  &  6.19   & 0.00		& 0.00 	& 	0.00 	&  0.00		& 0.66    \\
 		III    &  6.24  &  2.37 	&     0.02  	&	 0.28 	& 2.67 		&  0   \\
 		\hline \hline
 	\end{tabular}
 	\label{tab:theory-VNRA}
 \end{table}

 \section{Experimental Results}
 \subsection{Crystal Structure} 
 \begin{figure}[t]
 \renewcommand{\thefigure}{S\arabic{figure}}
\centering
\includegraphics[width=0.95\linewidth]{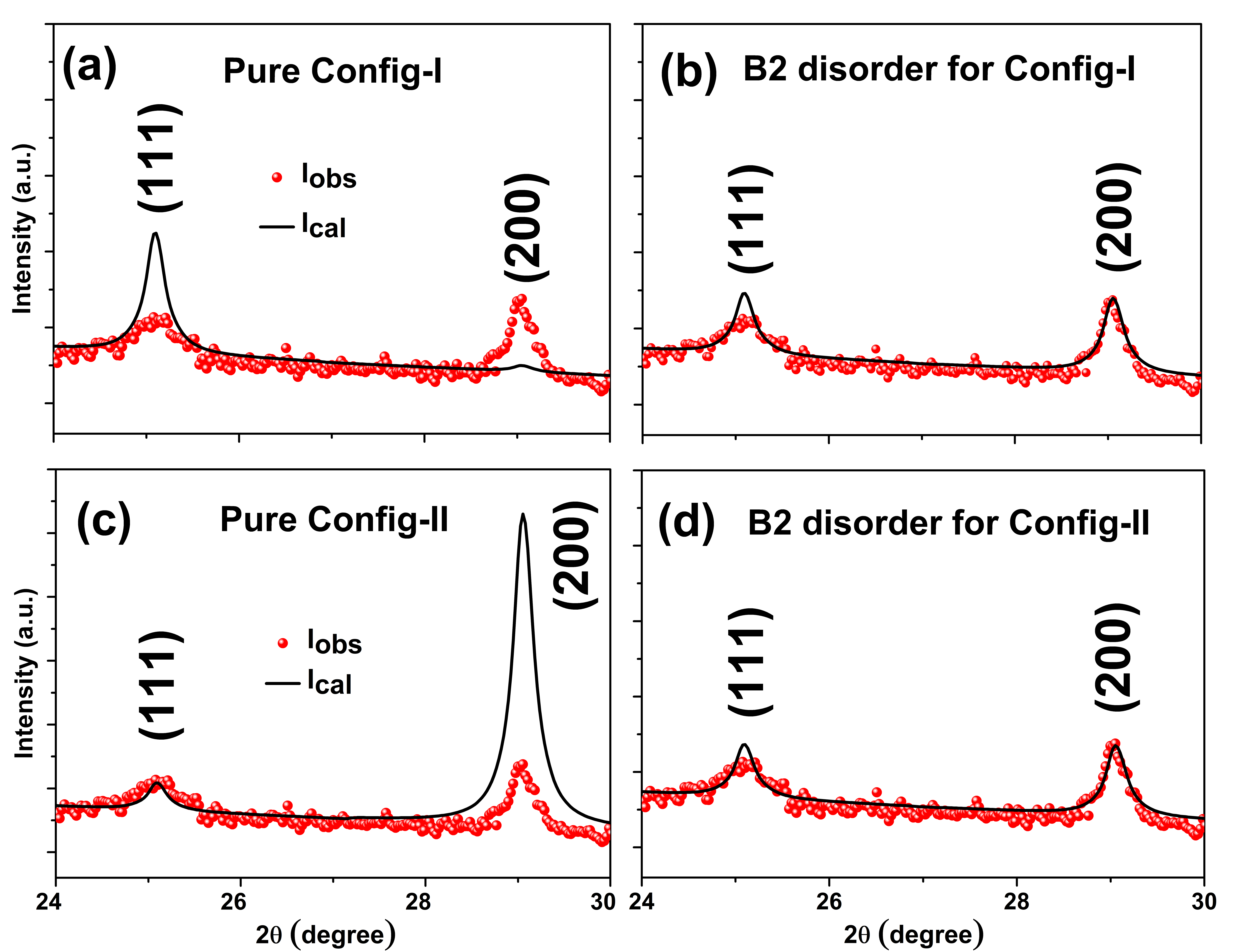}
\caption{Zoomed-in view near (111) and (200) peaks of XRD patterns for (a) pure configuration-I (with no disorder) (b) configuration-I with B$2$-disorder, (c) pure configuration-II (with no disorder) (d) configuration-II with B$2$ disorder, for VNbRuAl}
\label{fig:xrd-VNRA}
\end{figure}
 
 For more detailed analysis of XRD of VNbRuAl, the structure factor with the configuration I (Z-atom at 4$a$-site, X at 4$c$, X$'$ at 4$d$, and Y at 4$b$ site) can be expressed as 
\begin{equation}
\renewcommand{\theequation}{S\arabic{equation}}
F_{hkl} = 4(f_Z + f{_Y}e^{{\pi}i(h+k+l)} + f{_X}e^{\frac{{\pi}i}{2}(h+k+l)} + f_{X'}e^{-\frac{{\pi}i}{2}(h+k+l)}).
\label{eq:sfactor}
\end{equation}
with various $(hkl)$ values. Here $f_X$, $f_{X'}$, $f_Y$, and $f_Z$ are the atomic scattering factors for $X$, $X'$, $Y$, and $Z$ atoms respectively. For super lattice reflections i.e. for (111) and (200) planes the structure factor takes the forms,
\begin{equation}
\renewcommand{\theequation}{S\arabic{equation}}
F_{111} = 4[( f{_Y} - f_Z ) - i( f{_X} - f_{X'})]
\label{eq:sfactor111}
\end{equation}
\begin{equation}
\renewcommand{\theequation}{S\arabic{equation}}
F_{200} = 4[( f{_Y} + f_Z ) - ( f{_X} - f_{X'})]
\label{eq:sfactor200}
\end{equation}

Here, we first show the Rietveld refined fitting data with respect to the pure configuration I and II (without any disorder), as shown in the Fig. \ref{fig:xrd-VNRA}(a),(c). Clearly, these did not fit well  near (111) and (200) peaks for configurations  I and II respectively. Low intensity of (111) peak indicates the possibility of anti-site disorder at either the octahedral sites or the tetrahedral sites. Refinement considering 50\% anti-site disorder among the tetrahedral sites in configurations I and II looks better, but not perfect indicating absence of L2$_1$ disorder. The best fit was obtained for the configuration III by considering anti-site disorder between V-Ru and Nb-Al site atoms. This disorder arises due to equal probability of the octahedral sites to be occupied by Ru and V atoms and tetrahedral sites to be occupied by Al and Nb atoms. \cite{graf2011simple} The crystal structure corresponding to the best fit can be seen as B$2$ disorder in this alloy, as shown in Fig. 3 of the main manuscript. Hence, we conclude that VNbRuAl crystallizes in configuration III with B$2$ disorder.

 \subsection{Hall Measurements} 
\begin{figure}[t]
\renewcommand{\thefigure}{S\arabic{figure}}
\centering
\includegraphics[width=0.9\linewidth, right]{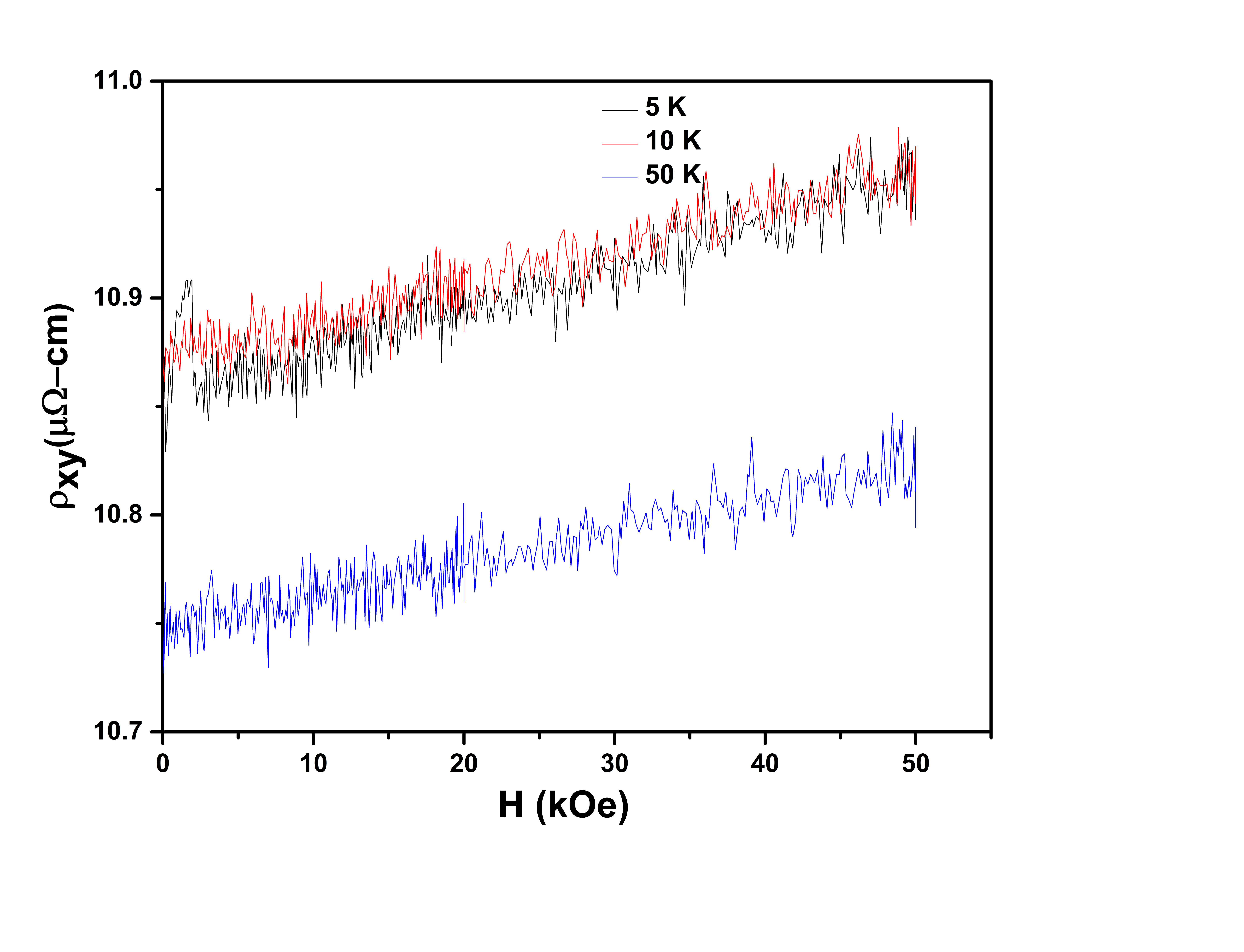} 
\caption{For VNbRuAl Hall resistivity ($\rho_{xy}$) vs. H at 5 K, 10 K and 50 K}.
\label{fig:Hall-VNRA}
\end{figure}

The Hall resistivity $ \rho_{xy}$ as a function of field was measured at various temperatures,  shown in Fig. \ref{fig:Hall-VNRA}. This shows an almost linear field dependence for VNbRuAl.



\end{document}